\documentclass[12pt]{article}

\usepackage[totalheight = 23cm, totalwidth = 17cm]{geometry}
\usepackage{amssymb,amsmath,amsfonts,amsbsy,epsfig}

\newcommand{\mpl}{M_{\rm Pl}}
\newcommand{\gfd}{g_{\rm 4D}}
\newcommand{\feff}{f_{\rm eff}}

\begin{document}

\begin{titlepage}

\begin{center}

\vspace*{-10ex}
\hspace*{\fill} SNUTP 07-013

\vskip 1.5cm

\Huge{Minimal gauge inflation}

\vskip 1cm

\large{
Jinn-Ouk Gong$^{1,2}$\footnote{jgong@hep.wisc.edu}
\hspace{0.2cm}\mbox{and}\hspace{0.2cm}
Seong Chan Park$^3$\footnote{spark@phya.snu.ac.kr}
\\
\vspace{0.5cm} {\em ${}^1$ Harish-Chandra Research Institute
\\
Chhatnag Road, Jhunsi, Allahabad 211 019, India
\\
\vspace{0.2cm} ${}^2$ Department of Physics, University of Wisconsin-Madison
\\
1150 University Avenue, Madison, WI 53706-1390, USA\footnote{Present address}
\\
\vspace{0.2cm} ${}^3$ Frontier Physics Research Division
\\
School of Physics and Astronomy, Seoul National University
\\
Seoul 151-747, Republic of Korea} }

\vskip 0.5cm

\today

\vskip 1.2cm

\end{center}

\begin{abstract}

We consider a gauge inflation model in the simplest orbifold $M_4 \times
S^1/\mathbb{Z}_2$ with the minimal non-Abelian $SU(2)$ hidden sector gauge symmetry.
The inflaton potential is fully radiatively generated solely by gauge
self-interactions. Following the virtue of gauge inflation idea, the inflaton, a
part of the five dimensional gauge boson, is automatically protected by the gauge
symmetry and its potential is stable against quantum corrections. We show that the
model perfectly fits the recent cosmological observations, including the recent WMAP
5-year data, in a wide range of the model parameters. In the perturbative regime of
gauge interactions ($\gfd \lesssim 1/(2\pi R\mpl)$) with the moderately compactified radius
($10 \lesssim R\mpl \lesssim 100$) the anticipated magnitude of the curvature perturbation
power spectrum and the value of the corresponding spectral index are in perfect
agreement with the recent observations.  The model also predicts a large fraction of
the gravitational waves, negligible non-Gaussianity, and high enough reheating
temperature.

\end{abstract}

\end{titlepage}

\setcounter{page}{0}
\newpage
\setcounter{page}{1}

\section{Introduction}

It is now widely accepted that an early period of accelerated expansion of the
universe, or inflation~\cite{inf}, can resolve many cosmological problems such as
horizon problem and can provide the desired initial conditions for the subsequent
hot big bang evolution of the observed universe~\cite{books}. Many observational
facts such as the flatness of the universe and the isotropy of the cosmic microwave
background (CMB) are natural consequences of inflation, and hence they strongly
support the existence of such a period of acceleration in the very early universe.
In particle physics point of view, inflation occurs when one or more scalar fields,
the inflaton fields, dominate the energy density of the universe with their
potential being overwhelming~\cite{Lyth:1998xn}. Under such a condition, dubbed
slow-roll condition, curvature perturbation $\mathcal{R}$ is produced which is
nearly scale invariant and is heavily constrained by the measurements of the
anisotropies of the CMB and the observations of the large scale
structure~\cite{obs}, including the recent Wilkinson Microwave Anisotropy Probe
(WMAP) 5-year data set~\cite{Komatsu:2008hk}. The slow-roll condition says that the
inflaton potential should be very flat, i.e. the effective mass of the inflaton
should be very small compared with the inflationary Hubble parameter. This is,
however, quite difficult to be achieved: for example, in supergravity, any generic
scalar field is expected to have an effective mass of
$\mathcal{O}(H)$~\cite{sugrainf}, which completely spoils the desired slow-roll
condition. It is thus natural to consider some symmetry principle which protects the
small inflaton mass from large radiative corrections or supergravity effects.

Shift symmetry, under which the field is invariant with respect to the
transformation
\begin{equation}
\phi \to \phi + a \, ,
\end{equation}
with $a$ being an arbitrary constant, is one of such symmetries and the
corresponding fields remain completely massless, i.e. their potential is exactly
flat as long as shift symmetry is unbroken. When the symmetry is explicitly broken,
they become pseudo Nambu-Goldstone bosons (pNGBs) and the potential acquires a tilt
depending on the symmetry breaking scale $f$ and is of the form
\begin{equation}\label{naturalinf_V}
V(\phi) = \Lambda^4 \left[ 1 \pm \cos\left( \frac{\phi}{f} \right) \right] \, .
\end{equation}
The model of natural inflation~\cite{naturalinf} makes use of such a pNGB with large
$f$, which renders the potential very flat. This largeness, however, requires $f \gg
\mpl \equiv G^{-1/2} \sim 10^{19} \mathrm{GeV}$ and hence the valid region of
successful natural inflation lies in the regime where we completely lose theoretical
control. This has been regarded as a significant drawback of natural inflation.

An idea to evade this problem was suggested in
Ref.~\cite{Arkani-Hamed:2003wu}\footnote{The idea of identifying the higher
dimensional gauge field as the inflaton is also suggested in
Ref.~\cite{Kaplan:2003aj}.}. In this scenario, called gauge inflation, the inflaton
field is essentially coming from a part of the fifth component of the gauge boson in
five dimensional bulk $A_5$, by which the gauge invariant Wilson line is defined as
\begin{equation}
e^{i\theta} \equiv \exp \left[i g \oint A_5 dy \right] \, ,
\end{equation}
where $g$ is the gauge coupling constant. By the one-loop interactions with the
charged particles, the potential of the canonically normalized field $\phi \equiv
f\theta$ has basically the same form as Eq.~(\ref{naturalinf_V}) while $f=1/(g_4 L)$
has extra dimensional nature with $L$ being the size of the fifth dimension. Hence
the potential can be trusted even when $f$ is larger than $\mpl$ in the perturbative
regime of gauge interaction, $g_4 \lesssim 1/(\mpl L)$. An interesting possibility
to unify the grand unification theory (GUT) and cosmic inflation in the framework of
the gauge inflation was suggested in Ref.~\cite{Park:2007sp} by one of the authors
of the present paper: starting from the higher dimensional $SU(5)$ GUT, the standard
model is recovered by orbifold projection by $S^1/{\mathbb{Z}_2}$ and a massless
scalar boson from the fifth component of the gauge boson is shown to play the role
of the inflaton with a fully radiatively generated flat potential.

In this paper, we take a different approach. Here we try to build the minimal model
of gauge inflation using the simplest orbifold $S^1/\mathbb{Z}_2$ and the minimal
non-Abelian gauge group $SU(2)$ for the hidden sector. The obvious advantages of
this approach is two-fold. First, by taking the hidden sector gauge interaction, the
theory is less constrained than the visible sector gauge theory, e.g.
GUT~\cite{Park:2007sp}. This is important since the required coupling constant is
typically very small, $g_4 \sim 10^{-2}$, so it is hard to reconcile this smallness
with the coupling constant unification. Second, the minimality can apply to the
particle contents. Different from the Abelian case, the non-Abelian gauge theory
accommodates the gauge self-interactions by which the potential for the inflaton
($\sim A_5$) can be generated even without any additional charged matter fields. So
the theory remains minimal in particle contents and its predictions are quite robust
and free from parameters such as the masses and other quantum numbers of newly
introduced matter fields. Combining the minimal choice of the orbifold and the gauge
group, no exotic particles appear in the theory. The model is clean.

The structure of the paper is as follows.
In the next section, we give the detail of the higher dimensional gauge theory and
outline the subsequent cosmological scenario. Section~\ref{sec_cos} is devoted to
the cosmological evolution of the model and we analytically calculate the observable
quantities produced during inflation and study their relations to the parameters of
the model. We also address the issue of reheating and estimate the reheating
temperature $T_\mathrm{RH}$. In Section~\ref{sec_con} we then conclude.

\section{The model}

As we already emphasized in the previous section, we do not wish to introduce any exotic particle in the hidden sector.
So non-Abelian gauge symmetry is introduced and the minimal choice is $SU(2)$.
The spacetime is five dimensional where the fifth dimension is compactified by an orbifold
$S^1/\mathbb{Z}_2$\footnote{If $1/R$ is $\mathcal{O}(\mathrm{TeV})$, the theory
could be relevant for the Higgs mechanism through the Hosotani
mechanism~\cite{Kubo:2001zc,Cacciapaglia:2005da}.}.

The $SU(2)$ gauge theory on the orbifold is constructed by specifying two independent parity conditions at the two
fixed points, $y=0$ and $y=\pi R$ where $R$ is the compactification radius, as
\begin{align}
A_\mu (x, -y) =& P_0 A_\mu (x, y) P_0 \, , \\
A_5 (x, -y) =& -P_0 A_5 (x, y) P_0 \, , \\
A_\mu (x, \pi R-y) =& P_1 A_\mu (x,\pi R+y) P_1 \, , \\
A_5 (x, \pi R-y) =& - P_1 A_5 (x, \pi R+y) P_1 \, ,
\end{align}
where $P_0$ and $P_1$ are $2\times 2$ matrices satisfying $P_0^2 = P_1^2 = 1$. The
translational transformation, $y \rightarrow y+2 \pi R$, is generated by successive
operation of the parity operators, $P_1 P_0$. Taking $P_0= P_1 =
\mathrm{diag}(1,-1)$, at the classical level the gauge symmetry $SU(2)$ is reduced
to $U(1)$ by the orbifold projection. Here we explicitly write down the parity
assignment with $P_0$ and $P_1$ as
\begin{align}
A_\mu &= \left(
            \begin{array}{cc}
              (++) & (--) \\
              (--) & (++) \\
            \end{array}
          \right), \\
A_5 &= \left(
            \begin{array}{cc}
              (--) & (++) \\
              (++) & (--) \\
            \end{array}
          \right),
\end{align}
thus the zero modes are $A_\mu ^3$ and $A_5^{1,2}$ which correspond to the $H=U(1)$
gauge boson, i.e. mirror photon and the scalar field, respectively. The scalar field
$A_5^{1,2}$ may develop a vacuum expectation value which could be put into the form
of $A_5^a \sim (\phi, 0,0)$ due to the remaining $U(1)$ global symmetry. Taking into
account the effects from the gauge, ghost and the scalar self interaction, the
one-loop effective potential for the field $\phi$ is induced as
\begin{equation}\label{1looppotential}
V_{\rm 1-loop}(\phi) = -\frac{9}{(2\pi)^6 R^4}\sum_{n=1}^\infty \frac{\cos
(n\phi/\feff)}{n^5} \, ,
\end{equation}
where the effective decay constant
\begin{equation}\label{feff}
\feff \equiv \frac{1}{\sqrt{2\pi R}g} = \frac{1}{2\pi \gfd R}
\end{equation}
is introduced for canonical normalization of $\phi$ \cite{Kubo:2001zc} (also see
\cite{Haba:2003ux}). Here we can add a cosmological constant
\begin{equation}\label{cc}
\frac{9\zeta(5)}{(2\pi)^6 R^4} \, ,
\end{equation}
where $\zeta(5) = \sum_{n=1}^\infty n^{-5}$, so that $V_\mathrm{1-loop}(0) = 0$ to
solve the cosmological constant problem, which we do not attempt to address. Thus,
the radiatively generated inflaton potential is the sum of
Eqs.~(\ref{1looppotential}) and (\ref{cc}), i.e.
\begin{equation}\label{potential}
V(\phi) = \frac{9}{(2\pi)^6R^4} \sum_{n=1}^\infty \frac{1}{n^5} \left[ 1 - \cos
\left( \frac{n\phi}{\feff} \right) \right] \, .
\end{equation}
In principle, we can further introduce additional matter field(s) with charge $q$
in the hidden sector but for the simplicity and predictability we introduce none. In
this sense, our model, which may be responsible for the cosmological inflation, is
the minimal model of hidden sector gauge theory on the simplest orbifold in five
dimensions. After an inflationary epoch achieved while $\phi$ rolls down the
effective potential given by Eq.~(\ref{potential}), it oscillates at the minimum and
decays to reheat the universe. After then the standard hot big bang evolution
follows. In the next section we calculate the observable quantities produced during
inflation and confirm the relevance of our model.

\section{Cosmological evolution}
\label{sec_cos}

\begin{figure}[h]
   \begin{center}
      \epsfig{file = 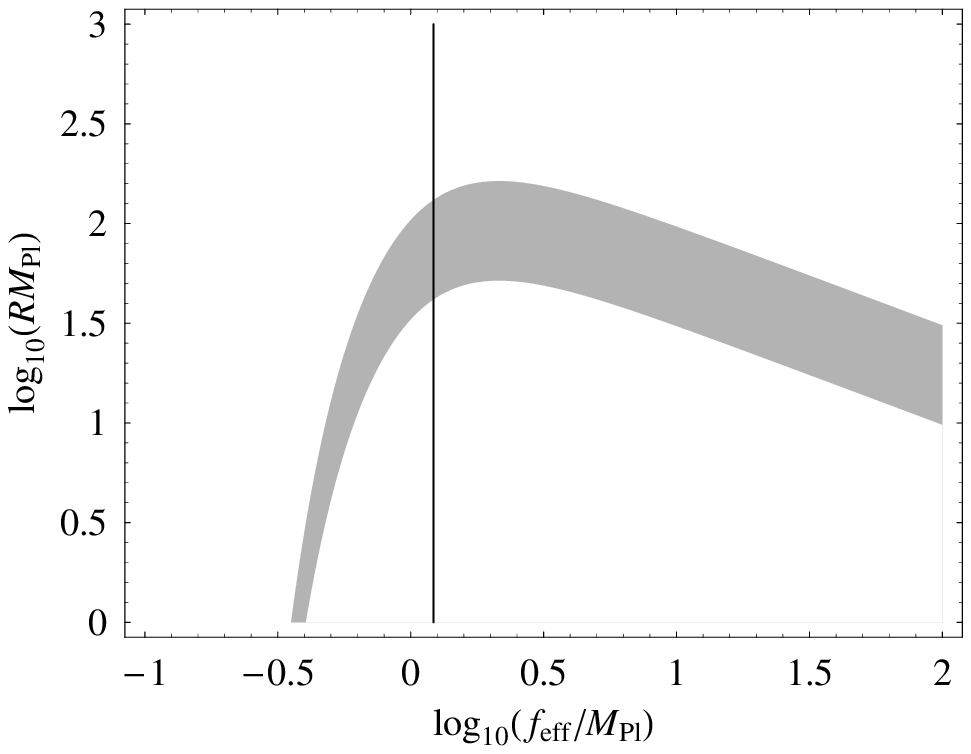, width = 8cm}%
      \epsfig{file = 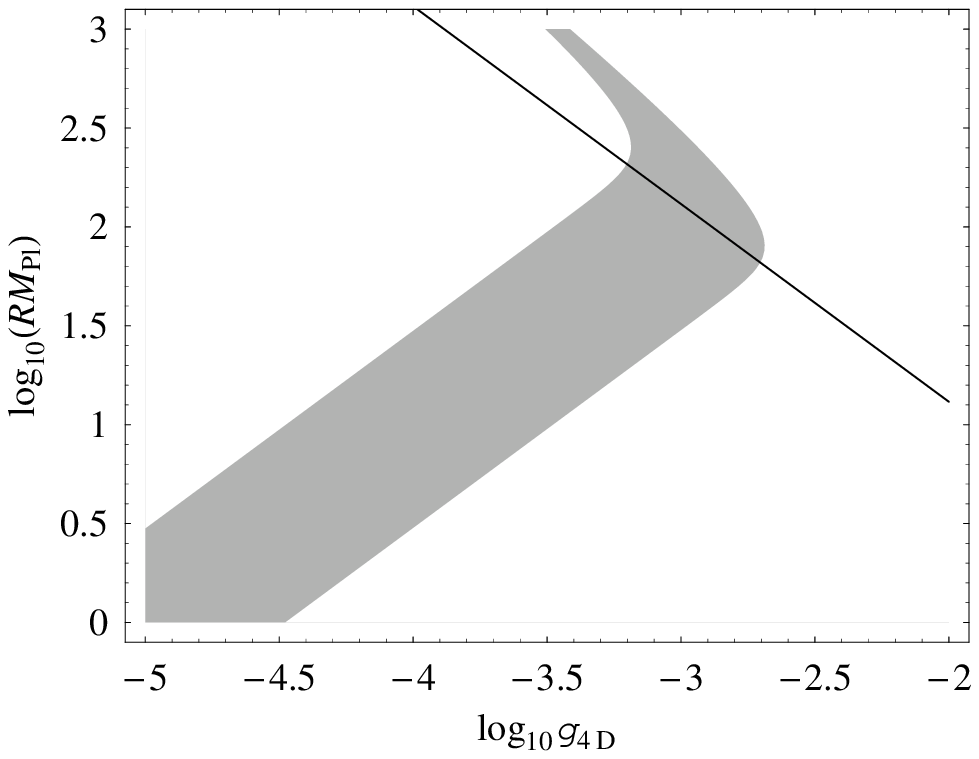, width = 8cm}%
   \end{center}
   \caption{The plot of $\mathcal{P}_\mathcal{R}^{1/2}$ in the (left) $\feff/\mpl - R\mpl$
   and (right) $\gfd  - R\mpl$ planes evaluated at $N = 60$. The shaded regions denote
   $10^{-5} \lesssim \mathcal{P}_\mathcal{R}^{1/2} \lesssim 10^{-4}$, and the solid
   lines correspond to $n_\mathcal{R} = 0.96$. Note that while for a large region
   $n_\mathcal{R}$ is saturated at $n_\mathcal{R} \approx 0.967$ (see Eq.~(\ref{n_largef})),
   only a limited region is allowed for $\mathcal{P}_\mathcal{R}^{1/2}$.}
   \label{Pandindex}
\end{figure}

In this section, we study the detail of the cosmological evolution of the model
described in the previous section. For analytic simplicity, we take as the leading
approximation only the first term of the sum in Eq.~(\ref{potential}), i.e.
\begin{equation}\label{approx}
V(\phi) \approx \frac{9}{(2\pi)^6 R^4} \left[ 1 - \cos \left( \frac{\phi}{\feff}
\right) \right] \, .
\end{equation}
As we will see in Table~\ref{comparison}, this approximation works fairly good. Then
the potential is identical to that of natural inflation, Eq.~(\ref{naturalinf_V}),
and the analytic calculations are straightforward especially when $\phi$ is close to
the top~\cite{Gong:2001he}. Here we just give the results of the observable
quantities: the power spectrum of the curvature perturbation
$\mathcal{P}_\mathcal{R}$, the corresponding spectral index $n_\mathcal{R}$, the
tensor-to-scalar ratio $r$ and the non-linear parameter
$f_\mathrm{NL}$~\cite{Komatsu:2001rj}. Under the slow-roll approximation, they are
given by
\begin{align}
\mathcal{P}_\mathcal{R}^{1/2} = & \sqrt{\frac{8V}{3\epsilon \mpl^4}} \, ,
\label{spectrum}\\
n_\mathcal{R} = & 1 - 6\epsilon + 2\eta \, ,
\label{index}\\
r = & 16\epsilon \, ,
\label{ttsratio}\\
-\frac{3}{5}f_\mathrm{NL} = & \frac{1}{2} \left[ \left( 3 + f_k \right) \epsilon -
\eta \right] \, ,
\label{fNL}
\end{align}
respectively. Here, $\epsilon$ and $\eta$ are the usual slow-roll parameters and are
defined by
\begin{align}
\epsilon \equiv & \frac{\mpl^2}{16\pi} \left( \frac{V'}{V} \right)^2 \, ,
\\
\eta \equiv & \frac{\mpl^2}{8\pi} \frac{V''}{V} \, ,
\end{align}
where a prime denotes a derivative with respect to $\phi$, and $f_k$ is a constant
in the range of $0 \leq f_k \leq 5/6$, depending on the
momenta~\cite{Maldacena:2002vr}. Note that the running of $n_\mathcal{R}$, which in
the slow-roll approximation is written as
\begin{equation}
\frac{dn_\mathcal{R}}{d\log{k}} = -16\epsilon\eta + 24\epsilon^2 + 2\xi^2 \, ,
\end{equation}
with
\begin{equation}
\xi^2 \equiv \frac{\mpl^4}{64\pi^2} \frac{V'V'''}{V^2}
\end{equation}
being another slow-roll parameter, is second order in the slow-roll approximation
and is negligibly small compared with the other quantities\footnote{In more general
classes of inflation models~\cite{generalSR}, we may obtain large enough
$dn_\mathcal{R}/d\log{k}$.}, so here we do not compute it while the calculation is
straightforward. Also, the running of $r$~\cite{Gong:2007ha},
\begin{equation}
\frac{d\log{r}}{d\log{k}} = 2(2\epsilon - \eta) \, ,
\end{equation}
is another first order quantity and thus could be observable, but we can obtain the
same result by combining Eqs.~(\ref{index}) and (\ref{ttsratio}) although it can
serve as a consistency check. Writing Eqs.~(\ref{spectrum}), (\ref{index}),
(\ref{ttsratio}) and (\ref{fNL}) in terms of $\feff$ and $R$ as
Eq.~(\ref{potential}), we can obtain
\begin{align}
\mathcal{P}_\mathcal{R}^{1/2} = & \frac{8\sqrt{3}}{(2\pi)^{5/2}}
\frac{\feff/\mpl}{(R\mpl)^2} \left\{ 2 -
\frac{32\pi(\feff/\mpl)^2}{16\pi(\feff/\mpl)^2 + 1} \exp \left[
\frac{-N}{8\pi(\feff/\mpl)^2} \right] \right\}
\nonumber\\
& \times \left\{ \frac{32\pi(\feff/\mpl)^2}{16\pi(\feff/\mpl)^2 + 1} \exp \left[
\frac{-N}{8\pi(\feff/\mpl)^2} \right] \right\}^{1/2} \, ,
\label{P_fR}
\\
n_\mathcal{R} = & 1 - \frac{1}{8\pi(\feff/\mpl)^2} \left\{ 2 +
\frac{32\pi(\feff/\mpl)^2}{16\pi(\feff/\mpl)^2 + 1} \exp \left[
\frac{-N}{8\pi(\feff/\mpl)^2} \right] \right\}
\nonumber\\
& \times \left\{ 2 - \frac{32\pi(\feff/\mpl)^2}{16\pi(\feff/\mpl)^2 + 1} \exp \left[
\frac{-N}{8\pi(\feff/\mpl)^2} \right] \right\}^{-1} \, ,
\label{n_fR}
\\
r = & \frac{1}{\pi(\feff/\mpl)^2} \frac{32\pi(\feff/\mpl)^2}{16\pi(\feff/\mpl)^2 +
1} \exp \left[ \frac{-N}{8\pi(\feff/\mpl)^2} \right]
\nonumber\\
& \times \left\{ 2 - \frac{32\pi(\feff/\mpl)^2}{16\pi(\feff/\mpl)^2 + 1} \exp \left[
\frac{-N}{8\pi(\feff/\mpl)^2} \right] \right\}^{-1} \, ,
\label{r_fR}
\\
-\frac{3}{5}f_\mathrm{NL} = & \frac{1}{16\pi(\feff/\mpl)^2} \left\{ 1 + \frac{1 +
f_k}{2} \frac{32\pi(\feff/\mpl)^2}{16\pi(\feff/\mpl)^2 + 1} \exp \left[
\frac{-N}{8\pi(\feff/\mpl)^2} \right] \right\}
\nonumber\\
& \times \left\{ 2 - \frac{32\pi(\feff/\mpl)^2}{16\pi(\feff/\mpl)^2 + 1} \exp \left[
\frac{-N}{8\pi(\feff/\mpl)^2} \right] \right\}^{-1} \, ,
\label{f_fR}
\end{align}
where
\begin{equation}
N \equiv \int H dt
\end{equation}
is the number of $e$-folds. Using Eq.~(\ref{feff}), we can write them in terms of
$\gfd$ and $R$ as
\begin{align}
\mathcal{P}_\mathcal{R}^{1/2} = & \frac{8\sqrt{3}}{(2\pi)^{7/2}\gfd(R\mpl)^3}
\left\{ 2 - \frac{8}{\pi(\gfd R\mpl)^2 + 4} \exp \left[ -N \frac{\pi}{2}(\gfd
R\mpl)^2 \right] \right\}
\nonumber\\
& \times \left\{ \frac{8}{\pi(\gfd R\mpl)^2 + 4} \exp \left[ -N \frac{\pi}{2}(\gfd
R\mpl)^2 \right] \right\}^{1/2} \, ,
\label{P_gR}
\\
n_\mathcal{R} = & 1 - \frac{\pi}{2} (\gfd R\mpl)^2 \left\{ 2 + \frac{8}{\pi(\gfd
R\mpl)^2 + 4} \exp \left[ -N \frac{\pi}{2} (\gfd R\mpl)^2 \right] \right\}
\nonumber\\
& \times \left\{ 2 - \frac{8}{\pi(\gfd R\mpl)^2 + 4} \exp \left[ -N \frac{\pi}{2}
(\gfd R\mpl)^2 \right] \right\}^{-1} \, ,
\label{n_gR}
\\
r = & \frac{32\pi(\gfd R\mpl)^2}{\pi(\gfd R\mpl)^2 + 4} \exp \left[ -N \frac{\pi}{2}
(\gfd R\mpl)^2 \right]
\nonumber\\
& \times \left\{ 2 - \frac{8}{\pi(\gfd R\mpl)^2 + 4} \exp \left[ -N \frac{\pi}{2}
(\gfd R\mpl)^2 \right] \right\}^{-1} \, ,
\label{r_gR}
\\
-\frac{3}{5}f_\mathrm{NL} = & \frac{\pi}{4}(\gfd R\mpl)^2 \left\{ 1 + \frac{1 +
f_k}{2} \frac{8}{\pi(\gfd R\mpl)^2 + 4} \exp \left[ -N \frac{\pi}{2} (\gfd R\mpl)^2
\right] \right\}
\nonumber\\
& \times \left\{ 2 - \frac{8}{\pi(\gfd R\mpl)^2 + 4} \exp \left[ -N \frac{\pi}{2}
(\gfd R\mpl)^2 \right] \right\}^{-1} \, .
\label{f_gR}
\end{align}
In Fig.~\ref{Pandindex} we show $\mathcal{P}_\mathcal{R}^{1/2}$ and $n_\mathcal{R}$
evaluated at $N = 60$ as functions of $\feff$, $R$ and $\gfd$. We also compare
analytic estimates with numerical results in Table~\ref{comparison}. As can be seen
from the table, Eq.~(\ref{approx}) is indeed a good enough approximation.

\begin{table}[h]
\begin{center}
    \begin{tabular}{l|l||*{2}{c|}c}
        \multicolumn{2}{c||}{} & $\mathcal{P}_\mathcal{R}^{1/2}$ & $n_\mathcal{R}$ & $r$
        \\
        \hline\hline
        $\log_{10}(\feff/\mpl) = 0.00$ & analytic & $4.96 \times 10^{-5}$ & 0.952 & 0.032
        \\
        \cline{2-5}
        $\log_{10}(R\mpl) = 2.04$ & numerical & $4.84 \times 10^{-5}$ & 0.955 &
        0.033
        \\
        \hline
        $\log_{10}(\feff/\mpl) = 0.50$ & analytic & $1.25 \times 10^{-5}$ & 0.967 &
        0.117
        \\
        \cline{2-5}
        $\log_{10}(R\mpl) = 2.04$ & numerical & $1.33 \times 10^{-5}$ & 0.967 &
        0.112
        \\
        \hline
        $\log_{10}(\feff/\mpl) = 1.00$ & analytic & $3.94 \times 10^{-5}$ & 0.967 &
        0.131
        \\
        \cline{2-5}
        $\log_{10}(R\mpl) = 1.54$ & numerical & $4.25 \times 10^{-5}$ & 0.967 &
        0.130
        \\
        \hline
        $\log_{10}(\feff/\mpl) = 1.50$ & analytic & $1.25\times 10^{-5}$ & 0.967 &
        0.131
        \\
        \cline{2-5}
        $\log_{10}(R\mpl) = 1.54$ & numerical & $1.33 \times 10^{-5}$ & 0.967&
        0.112
        \\
        \hline
        $\log_{10}(\feff/\mpl) = 2.00$ & analytic & $3.94 \times 10^{-5}$ & 0.967 &
        0.132
        \\
        \cline{2-5}
        $\log_{10}(R\mpl) = 1.04$ & numerical & $4.26\times 10^{-5}$ & 0.967 &
        0.134
    \end{tabular}
\end{center}
    \caption{From the top row, $R$ is chosen to make the inflationary energy scale
    $\Lambda = 10^{-3}\mpl$, $10^{-5/2}\mpl$ and $10^{-2}\mpl$. Also note that $r$ is
    fairly close to the observational sensitivity of near future experiments. As can be
    seen from this table, the leading approximation of taking $n = 1$ piece of
    Eq.~(\ref{potential}) is reasonably good.}
    \label{comparison}
\end{table}

From Eqs.~(\ref{n_fR}), (\ref{r_fR}) and (\ref{f_fR}), we can see that
$n_\mathcal{R}$, $r$ and $f_\mathrm{NL}$ are dependent only on the effective decay
constant $\feff$. This leads to the following simple expressions in the limit
$\feff/\mpl \to \infty$, which is favored for long enough inflation\footnote{In the
limit $\feff/\mpl \to \infty$, i.e. $\gfd \mpl \ll 1/(2\pi R)$, the gravitation
force, which scales as $(m^2/M_*^2)/r^{2+n}$ with $M_*$ being the cutoff mass scale
in $4+n$ dimensions, becomes stronger than the gauge force between two Kaluza-Klein
particles, $g^2/r^{2+n}$. In this parameter regime, the gravitational effects cannot
be neglected and the effective potential is apt to be modified: in this sense, the
naive idea of extranatural inflation is as unnatural as that of natural inflation.
See Ref.~\cite{ArkaniHamed:2006dz} for more detailed discussions.}, as
\begin{align}
n_\mathcal{R} \approx & 1 - \frac{4}{1 + 2N} \, , \label{n_largef}
\\
r \approx & \frac{16}{1 + 2N} \, ,
\label{r_largef}
\\
-\frac{3}{5}f_\mathrm{NL} \approx & \frac{2 + f_k}{2(1 + 2N)} \, ,
\label{f_largef}
\end{align}
respectively. Thus we can see that in this limit, evaluated at a certain $e$-folds
before the end of inflation, they have definite values independent of $\feff$ or
$R$. This is not surprising: huge $\feff$ means that the total number of $e$-folds
we can obtain is enormous, and the last 60 $e$-folds is only a final tiny fraction
of the whole expansion. Therefore the physical properties at this moment become
completely insensitive to the detail of the model, since already the inflationary
dynamics is following the late time attractor. It is this reason why we obtain
almost identical values of $n_\mathcal{R}$, $r$ and $f_\mathrm{NL}$ in the limit
$\feff/\mpl \to \infty$. This also means that the shape of $\mathcal{P}_\mathcal{R}$
is identical, meanwhile only its overall amplitude does depend on the inflationary
energy scale\footnote{See, e.g. Fig.~3 of Ref.~\cite{Gong:2006zp}.}.

\begin{figure}[h]
\begin{center}
   \epsfig{file = 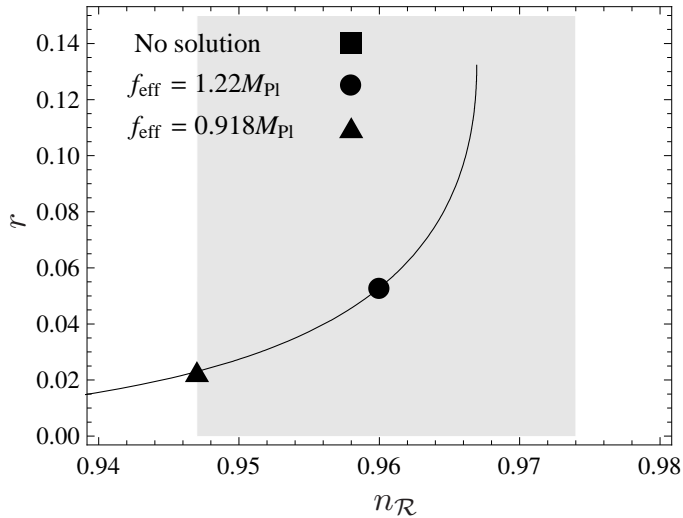, width = 9cm}%
\end{center}
   \caption{The prediction of the model in the $r-n_\mathcal{R}$ plane. Both $r$ and
   $n_\mathcal{R}$ are evaluated at 60 $e$-folds before the end of inflation.
   The shaded region shows the current observational $1\sigma$ bound derived from the
   WMAP 5-year + SN + BAO data: $n_\mathcal{R} = 0.960^{+0.014}_{-0.013}$. Our model
   is well within the observational upper abound and has a large range of parameter
   space satisfying the lower bound.}
   \label{r-ns}
\end{figure}

We show the $r- n_\mathcal{R}$ plot in Fig.~\ref{r-ns}. Note that as shown in
Eqs.~(\ref{n_largef}) and (\ref{r_largef}), they are saturated as $\feff/\mpl \to
\infty$, which corresponds to the upper right end of the curve where $n_\mathcal{R}
\approx 0.967$ and $r \approx 0.132$. The shaded region shows the current
observational $1\sigma$ bound $n_\mathcal{R} = 0.960^{+0.014}_{-0.013}$ which is
derived from the WMAP 5-year data combined with the observations of the type Ia
supernovae (SN) and the baryon acoustic oscillations (BAO)~\cite{Komatsu:2008hk},
and the points on the curve explicitly denote several constraints on
$n_\mathcal{R}$: the central value $n_\mathcal{R} = 0.960$ (circle), and the lower
bound $n_\mathcal{R}^{\rm lower} = 0.947$ (triangle). Our model is well below the
upper bound $n_\mathcal{R}^{\rm upper} = 0.974$ (square) and there is no solution
which corresponds to this point. The corresponding values of $r$ are 0.0528 and
0.0230, respectively. The current upper limit $r < 0.20$ (95\% confidence level)
encompasses the whole predicted range of $r$ of our model. For the observationally
allowed range of $n_\mathcal{R}$, $0.01 \lesssim r \lesssim 0.1$ and is large enough
to be detected within a few years by the forthcoming cosmological experiments and
therefore may serve as the first observational test. Also note that
$|f_\mathrm{NL}|$ is always much smaller than 1 and hence non-Gaussian signature is
absolutely not observable at all.

After inflation ends, the inflaton starts oscillation at the global minimum.
Although we are assuming no direct coupling between the hidden and the visible
sectors, they can communicate gravitationally and the energy stored in the inflaton
field can be converted to the light relativistic particles of the standard model to
reheat the universe. Let us estimate the reheating temperature $T_\mathrm{RH}$ via
the gravitational interaction in terms of the parameters of our model. With the
interaction rate
\begin{equation}\label{gravint}
\Gamma_\mathrm{grav} \sim \frac{m_\phi^3}{\mpl^2} \, ,
\end{equation}
using
\begin{equation}
m_\phi^2 \sim V'' \sim \frac{1}{\feff^2 R^4} =
\frac{\mpl^2}{(\feff/\mpl)^2(R\mpl)^4} \, ,
\end{equation}
we can write Eq.~(\ref{gravint}) as
\begin{equation}\label{gravint2}
\Gamma_\mathrm{grav} \sim \frac{\mpl}{(\feff/\mpl)^3(R\mpl)^6} \, .
\end{equation}
From the fact that inflation ends when $\dot\phi_\mathrm{end}^2 = V_\mathrm{end}$,
we can find the Hubble parameter at the end of inflation, under the approximation
Eq.~(\ref{approx}), as
\begin{align}\label{Hend}
H_\mathrm{end} = & \frac{3}{(2\pi)^{3/2}\pi} (R\mpl)^{-1} \left[ 16\pi(\feff/\mpl)^2
+ 1 \right]^{-1/2} R^{-1}
\nonumber\\
\sim & \mathcal{O}(0.1) \frac{R^{-1}}{(\feff/\mpl)R\mpl} \, .
\end{align}
Thus, for most parameter space $H_\mathrm{end} \gg \Gamma_\mathrm{grav}$ and the
energy transfer occurs well after inflation. We can now easily see that the
reheating temperature $T_\mathrm{RH}$ is estimated to be~\cite{lindebook}
\begin{equation}
T_\mathrm{RH} \lesssim \mathcal{O}(0.1) \sqrt{\Gamma_\mathrm{grav}\mpl} \sim
\mathcal{O}(0.1) \frac{\mpl}{(\feff/\mpl)^{3/2}(R\mpl)^3} \, .
\end{equation}
As an example, if we put $\feff/\mpl = 1$ and $R\mpl = 100$, the maximum reheating
temperature is estimated to be $T_\mathrm{RH} \sim 10^{12 - 13} \mathrm{GeV}$. The
universe then follows the well known hot big bang evolution\footnote{We may also
think of the inflaton decay through a messenger field at one-loop level even when
the inflaton field does not directly couple to the standard model fields. If a new
particle which is charged under the hidden as well as the standard model gauge
interactions exists, the inflaton field can couple to this new particle by the
hidden gauge interaction then through the standard model interaction the standard
model particles could be produced. The new particle can be the origin of the kinetic
mixing through the one loop interaction and plays the role of a messenger particle
as well. The contribution of the new particle to the inflaton potential can be still
negligible if the mass of the new particle is high enough as is assumed in the
paper.}.

\section{Conclusions}
\label{sec_con}

In this paper, we have presented a cosmological scenario from the hidden sector
$SU(2)$ gauge symmetry in the five dimensional orbifold $M_4\times
S^1/\mathbb{Z}_2$. The model is minimal in several aspects: the minimal non-Abelian
gauge group and the minimal orbifold compactification with the minimal number of
extra dimensions. Thanks to the non-Abelian nature, the bulk gauge boson, the fifth
component $A_5$ in particular, could have a one-loop induced effective potential
without introducing any exotic field in the model. This makes sure the minimality of
the model. The advantage of this minimal setup is as follows: the inflaton field is
a built-in ingredient of the theory and is automatically free from quantum
gravitational effects because of its higher dimensional locality and the gauge
symmetry. Fully radiatively generated one-loop potential is naturally able to
support a long enough period of slow-roll inflation provided that the theory is
weakly coupled, i.e. $\gfd \ll 1$, during the inflationary epoch. In very good
numerical precision, the minimal model essentially provides a realization of natural
inflation
\begin{eqnarray}
V(\phi) \approx \Lambda^4 \left[ 1-\cos \left( \frac{\phi}{\feff} \right) \right] \,
,
\end{eqnarray}
with $\Lambda^4 = 9R^{-4}/(2\pi)^6$ and $\feff = (2\pi\gfd R)^{-1}$. For $10
\lesssim R M_{\rm Pl} \lesssim 100$ and  $1 \lesssim \feff/\mpl \lesssim 100$, the
model predicts the observable cosmological quantities
\begin{equation}
1.2 \times 10^{-5} \lesssim \mathcal{P}_\mathcal{R} \lesssim 4.9 \times 10^{-5} \, ,
\end{equation}
\begin{equation}
0.952 \lesssim n_\mathcal{R} \lesssim 0.966 \, ,
\end{equation}
\begin{equation}
0.03 \lesssim r \lesssim 0.13 \, .
\end{equation}
The power spectrum of the curvature perturbation $\mathcal{P}_\mathcal{R}$ and the
corresponding spectral index $n_\mathcal{R}$ are in good agreement with the current
observations. While $|f_\mathrm{NL}|$ is always far smaller than 1 and no detectable
non-Gaussianity is expected, very interestingly the predicted tensor-to-scalar ratio
$r$ is quite close to sensitivity of the near future cosmological experiments. This
would be the first test of our minimal cosmological model. The reheating temperature
$T_\mathrm{RH}$ is estimated to be high enough to successfully follow the standard
hot big bang evolution.

\subsection*{Acknowledgements}

We are grateful to Misao Sasaki and the Yukawa Institute for Theoretical Physics at
Kyoto University where some part of this work was carried out during ``Scientific
Program on Gravity and Cosmology'' (YITP-T-07-01) and ``KIAS-YITP Joint Workshop:
String Phenomenology and Cosmology'' (YITP-T-07-10). JG thanks Daniel Chung, L.
Sriramkumar and Ewan Stewart for helpful conversations, and is partly supported by
the Korea Research Foundation Grant KRF-2007-357-C00014 funded by the Korean
Government. SCP appreciates Yasunori Nomura for his comments on low energy
constraints of $\gfd$ and also thanks C. S. Lim for encouragement to publish this
paper.

\end{document}